\documentclass[3p, twocolumn]{elsarticle}

\usepackage{hyperref}
\usepackage{subfigure}
\usepackage[usenames,dvipsnames]{xcolor}
\usepackage{amssymb}

\journal{...}

\graphicspath{{figs/}}
\newcommand{\mathnotation}[2]{\newcommand{#1}{\ensuremath{#2}}}
\mathnotation{\Sil}{\textrm{SiO}_2}
\mathnotation{\Ba}{\textrm{BaO}}
\mathnotation{\Bo}{\textrm{B}_2\textrm{O}_3}

\begin{document}

\begin{frontmatter}

\title{Hydrodynamic coarsening in phase-separated silicate melts}

\author{David Bouttes}
\address{Laboratoire PMMH, UMR 7636 CNRS/ESPCI/Univ. Paris 6
UPMC/Univ. Paris 7 Diderot, 10 rue Vauquelin, 75231 Paris cedex 05,
France}

\author{Oc\'eane Lambert}
\author{Corinne Claireaux}
\author{William Woelffel}
\author{Davy Dalmas}
\author{Emmanuelle Gouillart \footnote{corresponding author}}
\address{Surface du Verre et Interfaces, UMR 125 CNRS/Saint-Gobain,
93303 Aubervilliers, France}

\author{Pierre Lhuissier}
\author{Luc Salvo}
\address{SIMAP, GPM2 group, CNRS UMR 5266, University of Grenoble
38402 Saint Martin d'H\`eres
France}

\author{Elodie Boller}
\address{European Synchrotron Radiation Facility (ESRF), BP 220,
38043 Grenoble, France}

\author{Damien Vandembroucq}
\address{Laboratoire PMMH, UMR 7636 CNRS/ESPCI/Univ. Paris 6
UPMC/Univ. Paris 7 Diderot, 10 rue Vauquelin, 75231 Paris cedex 05,
France}

\fntext[myfootnote]{corresponding author}


\begin{abstract}

Using \emph{in-situ} synchrotron tomography, we investigate the
coarsening dynamics of barium borosilicate melts during phase separation.
The 3-D geometry of the two interconnected phases is determined thanks to
image processing. We observe a linear growth of the size of domains with
time, at odds with the sublinear diffusive growth usually observed in
phase-separating glasses or alloys. Such linear coarsening is attributed
to viscous flow inside the bicontinuous phases, and quantitative
measurements show that the growth rate is well explained by the ratio of
surface tension over viscosity. The geometry of the domains is shown to
be statistically similar at different times, provided that the
microstructure is rescaled by the average domain size. Complementary
experiments on melts with a droplet morphology demonstrate that viscous
flow prevails over diffusion in the large range of domain sizes measured
in our experiments (1 - 80 $\mu$m).

\end{abstract}


\end{frontmatter}

\section{Introduction}

Liquid-liquid phase separation is an appealing mechanism for building
materials with tailored microstructure~\cite{Levitz1991}. Molten
materials such as polymers~\cite{Lloyd1991}, silicate
melts~\cite{Craievich1983, Mazurin1985, Huang1991} or metallic
alloys~\cite{Davidoff2013, He2013, Kim2013} exhibit liquid-liquid
immiscibility under a critical temperature~\cite{Balluffi2005}. During a
quench under the immiscibility dome, spinodal decomposition or nucleation
and growth are first responsible for the growth of composition
fluctuations. Then the size of domains of fixed composition increases in
order to minimize interfacial energy -- a later stage called
\emph{coarsening}~\cite{Bray1994}. The ability to tune the typical size
of the microstructure by controlling coarsening kinetics is paramount for
applications of phase separation, among which porous
glasses~\cite{Wiltzius1987} and membranes~\cite{Kukizaki2010}, cellular
materials for the food industry~\cite{Tanaka2009,Tanaka2012},
core-shell particles~\cite{Shi2013}, quantum dots~\cite{Park2005} or
super-hydrophobic thin films~\cite{Nakajima2000}.

For most materials, coarsening results from molecular diffusion
through interfaces between the phases, transporting matter from
highly-curved interfaces to flatter interfaces. This mechanism operates
whatever the topology of the phases -- either for interconnected domains
or for droplets
~\cite{Lifshitz1961}. Diffusive coarsening is a slow transport mechanism,
with the typical domain size growing with time $t$ as $t^{1/3}$. Only
limited variations in domain sizes with heating time can therefore be
observed in experiments. Amongst inorganic materials, a large body of
the phase separation literature is devoted to silicate
melts~\cite{Mazurin1985}. The seminal theoretical work of
Cahn~\cite{Cahn1965} finds its inspiration in the geometry of
phase-separated sodium borosilicate glasses. Kinetic studies in silicates
evidenced diffusive coarsening both for a droplet~\cite{Wheaton2007,
Craievich1983} and
an interconnected microstructure~\cite{Dalmas2007}. Because of the
limited range of sizes that are observable during a $t^{1/3}$ evolution,
a more successful approach for controlling the microstructure size
consists in leveraging thermodynamics instead of kinetics, by changing
the composition in order to modify the value of the critical temperature.
Martel et al.~\cite{Martel2011}, for example, found that the typical size
of phase-separated droplets decreased exponentially with alumina content
in calcium aluminosilicate glasses. A similar strategy has been applied
to metallic glasses~\cite{Han2014}.

However, another coarsening mechanism has been proposed for larger domain
sizes, namely hydrodynamic coarsening, that can occur because of viscous
flow driven by Laplace pressure~\cite{Siggia1979}. This regime is
characterized by a linear growth of domain sizes with time, that should
be much more flexible to tune the microstructure, for a given
composition. Several numerical studies~\cite{Kendon2001, Ahmad2010} have
addressed hydrodynamic coarsening; however, experimental observations of
this regime are confined to the field of colloid-polymer
mixtures~\cite{Aarts2005, Bailey2007} and polymers~\cite{Sung1996,
Yuan2005}, with no single experimental evidence in the field of inorganic
materials. Since mechanisms such as crystallization (in alloys and
glass-forming melts), gravity or convection and shear often disrupt
coarsening and result in different morphologies, one may wonder if a
linear growth of domains can be observed at all in inorganic materials.

New insights on phase separation paralleled the emergence of experimental
techniques in materials science. Whereas the first kinetic studies were
realized in the Fourier space, using light diffusion
techniques~\cite{Boiko1970, Levitz1991}, direct access to the morphology
of the phases is now possible thanks to imaging techniques such as
electronic microscopy~\cite{Elmer1970}, atomic force
microscopy~\cite{Dalmas2007, Wheaton2007}, X-ray
tomography~\cite{Momose2005} or atom probe tomography~\cite{Roussel2013,
Han2014}. In-situ synchrotron microtomography~\cite{Baruchel2006} is a
technique of choice for studying microstructure
formation~\cite{Limodin2007, Gouillart2012}, since it provides both the
3-D microstructure and its topology, and successive snapshots of its
evolution. Recently~\cite{Bouttes2014}, our team started using in-situ
tomography to study phase separation in silicate melts. 

In this work, we investigate the coarsening stage of a barium
borosilicate glass-forming melt, far above its glass transition. Thanks
to in-situ tomography, we evidence a linear growth of the microstructure
with time, characteristic of viscous coarsening, that is shown to stem
from a series of hydrodynamic pinch-offs. In Section~\ref{sec:matmet}, we
present the materials and methods used in this work, in particular the
characterization of the glasses and the in-situ tomography experiments.
Section~\ref{sec:coarsening} focuses on the coarsening kinetics during
isothermal treatments observed by in-situ microtomography,
for an interconnected geometry. Section~\ref{sec:scaling} studies the
evolution of the 3-D geometry of the phases during the coarsening
process. Finally, we discuss in Section~\ref{sec:discussion} the relative
importance of diffusive and viscous transport in our system.

\section{Materials and methods \label{sec:matmet}}

\subsection{Glass composition and preparation}

\begin{figure}
\centerline{\includegraphics[width=0.99\columnwidth]{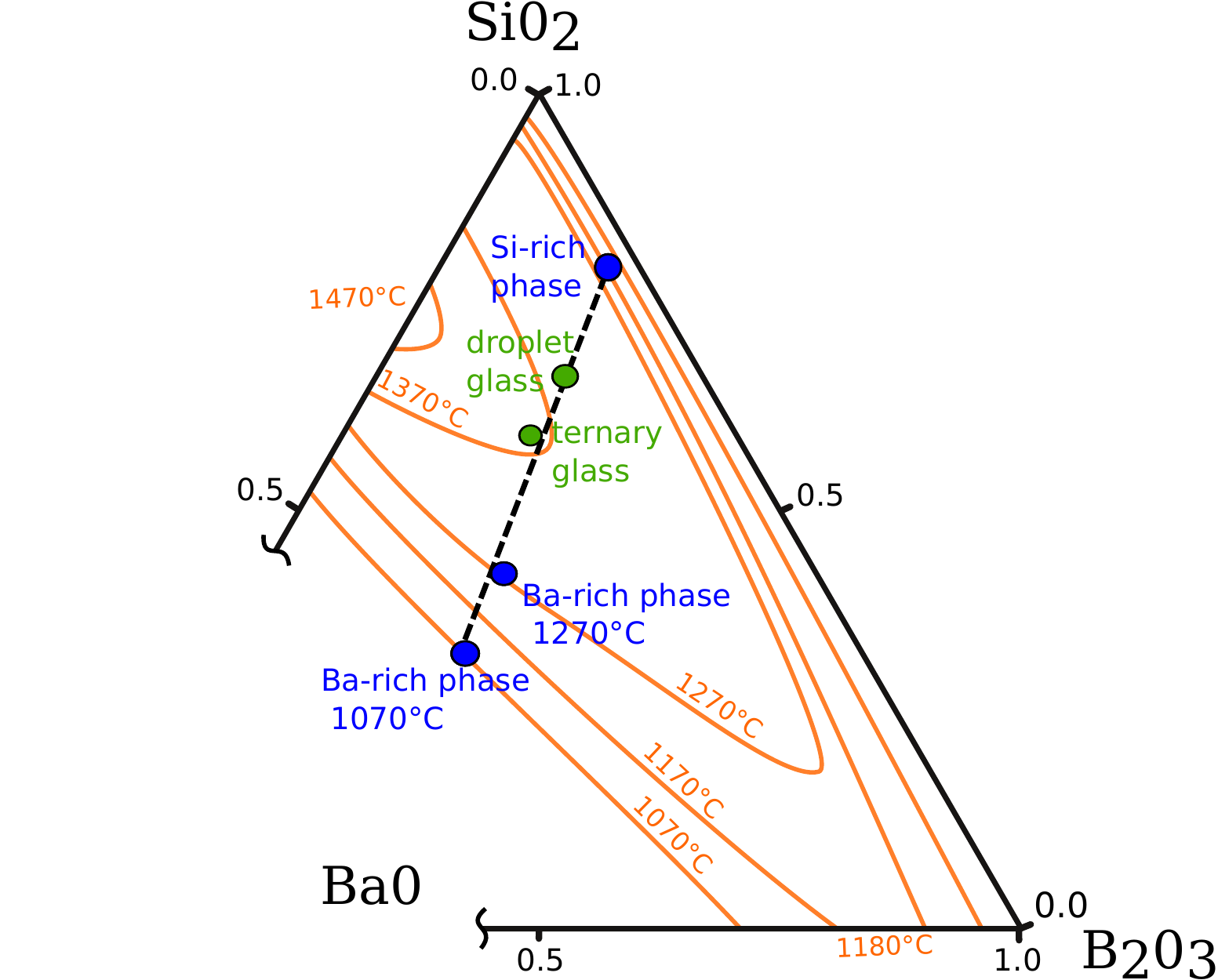}}
  \caption{Phase diagram of the $\Sil - \Ba - \Bo$ system, in weight
fractions. Isotherms of
the immiscibility dome (in orange) are plotted after~\cite{Levin1958},
while the tie-line corresponding to composition 1 is drawn from electron
microprobe measurements. The composition used in tomography experiments
corresponds to the \emph{ternary glass} point.\label{fig:diagram}}
\end{figure}

In this work, we focus on the ternary diagram $\Sil - \Ba - \Bo$. While
alkali borosilicates have been much studied in the literature,
alkaline-earth borosilicates are known to have a higher critical
temperature, and a larger immiscibility domain~\cite{Mazurin1985}. The
immiscibility diagram determined by Levin and Cleek~\cite{Levin1958} is
shown in Fig.~\ref{fig:diagram}. The critical temperature of the system
is above $1470^\circ$C, and a large fraction of the immiscibility dome is
located above liquidus surfaces (not shown here), resulting in stable
liquid-liquid phase separation. 

\begin{table*}
\begin{center}
\begin{tabular}{|c|c|c|c|c|}
\hline
glass name &  $\Sil$ & $\Ba$ & $\Bo$ & $\textrm{Al}_2\textrm{O}_3$ \\
\hline 
ternary glass & 59 & 21.5 & 19.5 & 0 \\
Si-rich phase & 80 & 2.7 & 17.3 & 0\\
1270-Ba-rich phase & 43 & 32 & 25 & 0\\
1070-Ba-rich phase & 32.8 & 41.1 & 26.1 & 0\\
quaternary glass & 57 & 21 & 20 & 2 \\
droplet glass & 66 & 14 & 20 & 0\\
\hline
\end{tabular}
\end{center}
\caption{Composition (in weight percent) of the different barium
borosilicate glasses. \label{tab:tab}}
\end{table*}

The ternary composition studied here is found in Table \ref{tab:tab}. It
corresponds to a composition well
inside the immiscibility domain in order to obtain an interconnected
microstructure, and to a liquidus temperature of $1180^\circ$C.

In order to study submicronic microstructures, we have studied a second
glass composition with a small percentage of
alumina~\cite{Du2000,Martel2011}, denoted as \emph{quaternary glass} in
Table \ref{tab:tab} and subsequently. Since the quaternary diagram has
not yet been characterized in the literature, we determined the dome
temperature for this composition by thermal treatments at different
temperatures, followed by a rapid quench in water. The dome temperature
was estimated to be $1175^\circ$C ($\pm 25^\circ$C).

For the elaboration of the glasses, batches of 1 kg were melted at
$1600^\circ$C from raw materials (reagent-grade barium carbonate and
boric acid, and E400 silica) in a gas-fired lab furnace. Oxygen bubbling
was used to ensure a good homogeneity of the melt. Glasses were
quenched in air and annealed at $650^\circ$C. During the quench from
$1600^\circ$ C to room temperature, a small microstructure appeared
because of phase separation. The typical size of this initial
microstructure is $5\, \mu \textrm{m}$ for the ternary glass, and $100$
nm for the quaternary glass (because of a much lower dome temperature).

\subsection{Tie-line}
 
Determining phase compositions in demixed glasses is usually a difficult
endeavour because of the submicronic size of the domains. Typical methods
use selective dissolution in acid~\cite{Charles1964} or compare DSC plots
to previously-determined glass-transition isotherms~\cite{Mazurin1972},
but such methods often lack precision. In our system, large sizes of the
domains (several tens of microns) can easily be obtained, which made it
possible to analyze the compositions by electron microprobe. Thermal
treatments were realized at $1070^\circ$C and $1270^\circ$C for the
ternary glass. The composition of the phases was analyzed using a Cameca
SX 100 microprobe at 15 kV and 15 nA. The beam was unfocused with a spot
size of $20\,\mu \textrm{m}$ in order to limit boron migration. The
compositions of the phases are given in Table \ref{tab:tab} and
represented in Fig.~\ref{fig:diagram}. Compositions are in agreement with
the limits of the immiscibility region proposed by Levin and
Cleek~\cite{Levin1958}. At both temperatures, the glass melt separates in
a silica-rich phase and a barium-rich one, the variation of the boron
content being much smaller. The precision of the microprobe measurements
-- of the order of 1 wt\% because of the incertitude on boron content --
is not sufficient to distinguish between the silica-rich compositions at
$1070^\circ$C and $1270^\circ$C. Interestingly, phases at $1070^\circ$C
and $1270^\circ$C lie on the same tie-line, therefore no tie-line
rotation is observed in such a temperature range (contrary to other
studies made at lower temperature~\cite{Scholes1970}, closer to the glass
transition).

\subsection{Physical properties}

As will be described later, the kinetics of domain coarsening is governed
by the ratio between the interfacial tension between the two phases, and
the viscosity of the most viscous phase. In order to characterize the
physical properties of the separated phases, glasses with the composition
of the silica-rich phase, and the barium-rich phases at $1070^\circ$C and
$1270^\circ$C, were melted from raw materials using the same procedure as
for the ternary glass.  

\begin{figure}
\begin{minipage}{0.33\columnwidth}
\subfigure[]{
\centerline{\includegraphics[width=0.99\textwidth]{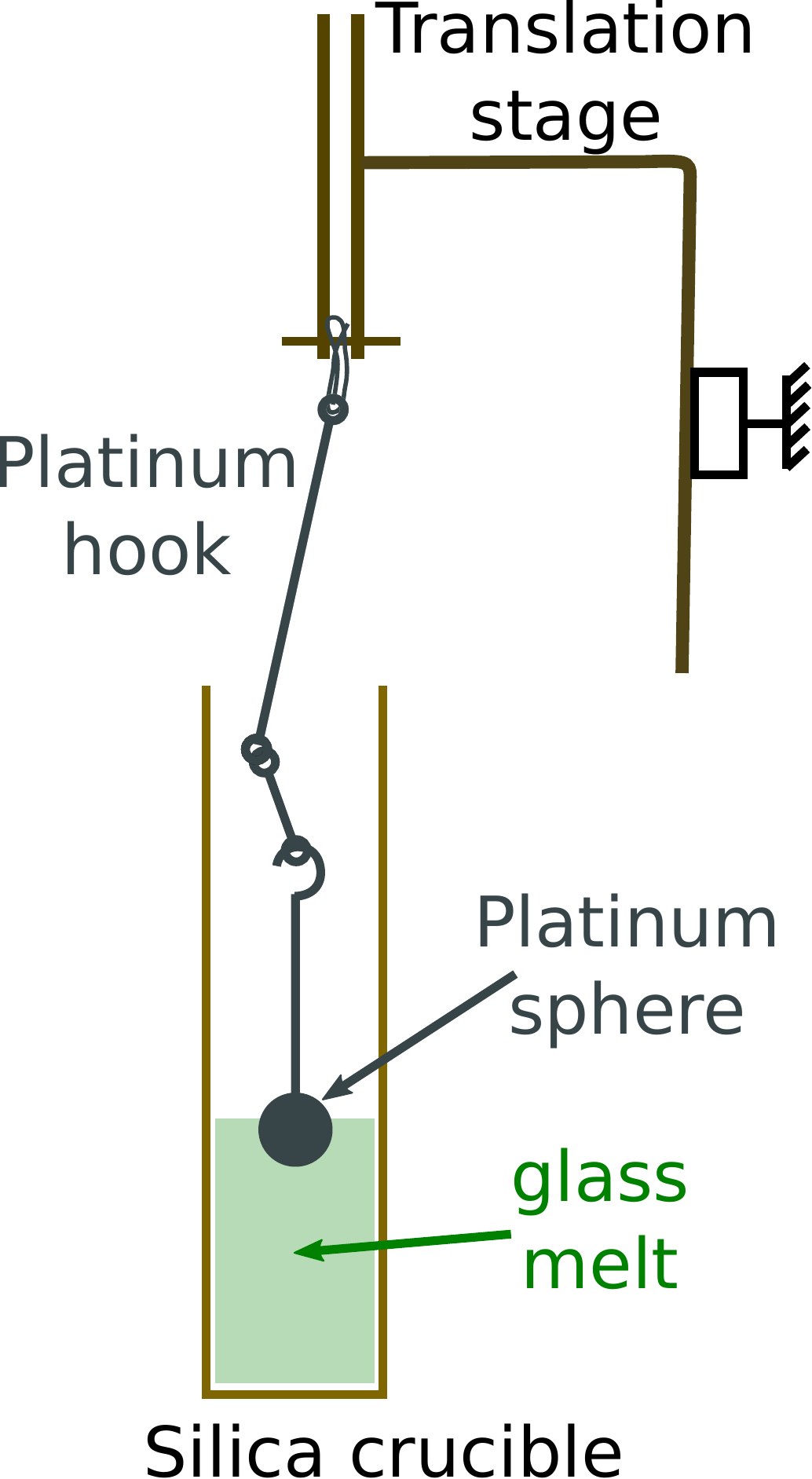}}
}
\end{minipage}
\begin{minipage}{0.65\columnwidth}
\subfigure[]{
\centerline{\includegraphics[width=0.99\textwidth]{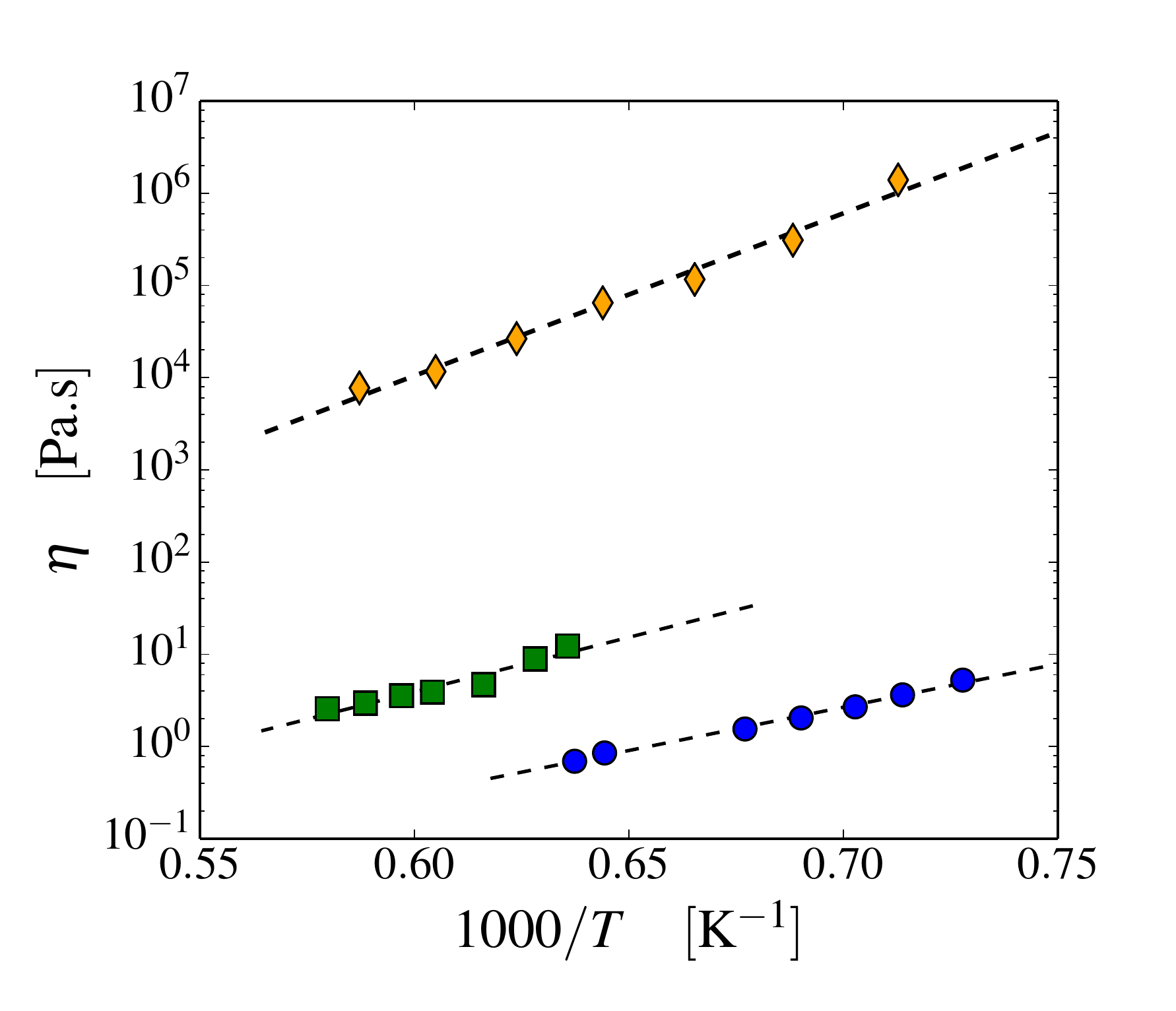}}
}
\end{minipage}
  \caption{(a) Schematic of the falling sphere experiment. The diameter
of the sphere is 1 cm, and the inner diameter of the crucible is 5 cm.
(b) Arrhenian plot of measured viscosities for the silica-rich phase
(\textcolor{YellowOrange}{$\blacklozenge$}), $1270^\circ$-barium-rich phase
(\textcolor{OliveGreen}{$\blacksquare$}) and $1070^\circ$-barium-rich
phase (\textcolor{blue}{$\bullet$}).
\label{fig:visco}}
\end{figure}

\paragraph{Viscosity} The viscosities of the barium-rich phases were
measured using a high-temperature Couette-flow apparatus. For the
silica-rich phase, the viscosity was too high to use the Couette set-up.
We therefore resorted to falling-sphere viscosimetry for the silica-rich
phase. Fig.~\ref{fig:visco} (a) shows a schematic of the experiment. A
transparent silica crucible containing the glass melt is placed inside an
electric furnace with a transparent silica window. A platinum sphere is
tracked with a camera while it falls through the melt, and Stokes' law is
used in order to retrieve the viscosity of the melt. Corrections were
applied to account for the finite size of the crucible~\cite{Brenner1961}
and the accuracy of the method was checked using a soda-lime glass of
known viscosity.

The viscosities of the different phases are represented on an Arrhenian
plot in Fig.~\ref{fig:visco} (b). A very important viscosity contrast is
observed between the silica-rich phase and the barium-rich phase, up to
five orders of magnitude. 

In order to supplement viscosity data, dilatometric measurement were
realized on the silica-rich phase and the $1070^\circ$C-barium-rich phase.
The glass transition and the dilatometric softening point (corresponding
to a viscosity of about $10^{11} \textrm{Pa}.\textrm{s}$) were obtained
with this method, with values of respectively 570 and $690^\circ$C for
the silica-rich phase, and 640 and $670^\circ$C for the
$1070^\circ$-barium-rich phase. Such measures confirm the large extent of
the temperature range at which phase separation occurs.

\paragraph{Interfacial tension} We estimated the value of the
interfacial tension $\gamma$ between the two phases to be of the order of
$10^{-2}\,\textrm{J.m}^{-2}$, from the comparison of experimental images
of the phases and numerical hydrodynamic simulations. The numerical
method is described in the Appendix. The order of magnitude is consistent
with the scarce literature on measurements of interfacial tension between
glassy melts~\cite{Dittmar2013, Wondraczek2005}, albeit at the lower end
of reported values.

\paragraph{Densities} The density of the separated phases were measured
from the bulk synthesized glasses, using Archimedes principle. With a
density difference of $\Delta\rho \simeq 10^3 \,\textrm{kg}.\textrm{m}^{-3}$
between the Si-rich and the 1070-Ba-rich phase, we estimate the Bond
number
\begin{equation}
Bo = \frac{\Delta\rho g l^2}{\gamma} \simeq 10^{-2}
\end{equation}
for a typical size $\ell$ of $100\, \mu\textrm{m}$, which is the maximum
size that is reached in our coarsening experiments.
The Bond number measures the ratio between gravitational and interfacial
forces. Its small value means that gravity can be neglected in comparison
to Laplace pressure.

\subsection{Imaging methods and thermal treatments}

In-situ X-ray microtomography experiments were realized at the ID19
beamline of the ESRF synchrotron. The imaged samples were 2-mm-diameter
cylinders machined from the bulk ternary glass and put inside
3-mm-outer-diameter alumina crucibles. On the beamline, we used the
dedicated ``Ecole des Mines" furnace, that can heat millimetric-sized
samples to temperatures up to $1500^\circ$C. During in-situ experiments,
X-rays can get through the furnace thanks to a thin silica window.
Samples are glued on top of an alumina rod fixed on a Leuven fast
rotation stage, and are allowed to rotate freely inside the fixed furnace
thanks to a hole at the bottom of the furnace. At the beginning of a
thermal treatment, the sample is introduced into the already hot furnace,
ensuring a fast heating rate: we measured that it only takes a few tens
of seconds for the sample to reach working temperatures above
$1000^\circ$C. Isothermal experiments were realized at temperatures
ranging from $1030^\circ$C to $1330^\circ$C.

Pink beam with an energy of 32 keV is used to achieve sufficient
transmission. At this energy, the absorption coefficients of the two
phases (resp. 0.08 $\textrm{mm}^{-1}$ and 1 $\textrm{mm}^{-1}$ for the
silica-rich and the 1070$^\circ$-barium-rich phases) are very
different, resulting in a good absorption contrast that facilitates
further image processing. The X-ray photons transmitted through the
sample were converted to visible light by a 25 $\mu \textrm{m}$-thick
LuAG scintillator. The pixel size is $1.1\, \mu \textrm{m}$. The PCO
Dimax camera is equipped with a fast CMOS detector; acquisition times for
one radiography varied from 6 to 30 ms depending on the working
temperature, resulting in total acquisition times from 4 to 10 s for the
total number of radiographies acquired over a rotation of $180^\circ$.

3-D absorption images were reconstructed from the set of radiographies
using a standard filtered back-projection algorithm~\cite{Mirone2014}.
Python's \texttt{scikit-image}~\cite{VanderWalt2014} was used for further
processing of 3-D images. Voxels were attributed to the two phases of
the glass using the following segmentation procedure: a first denoising
step is realized with a total-variation filter~\cite{Chambolle2004}; then
the Random Walker algorithm~\cite{Grady2006} is used to segment the two
phases. 3-D visualizations of the minority-phase surface are realized
with the Mayavi \cite{ramachandran2011mayavi} Python package.

For complementary experiments at scales out of the reach of parallel-beam
synchrotron tomography, thermal treatments were realized on
centimetric-sized samples in an electric furnace. The samples were
quenched and impregnated in an epoxy resin. Cross-sections were polished
and imaged with a Zeiss DSM 982 Gemini scanning electron microscope,
operated at 20 kV.

\section{Viscous coarsening \label{sec:coarsening}}

\begin{figure}
\centerline{\includegraphics[width=0.99\columnwidth]{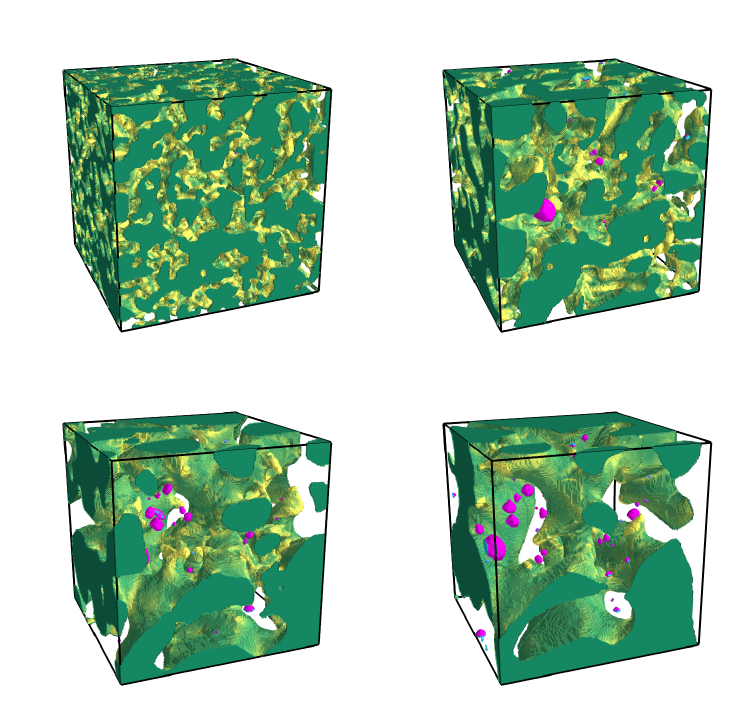}}
\caption{Tomography images of domain coarsening, for an experiment at
$1230^\circ$C and heating times of 80, 160, 320 and 480 s. The size of
the square box is 250 $\mu$m. Only the minority (barium-rich) phase is
represented. Shades of green encode the local mean curvature, from dark
green to yellow. Isolated domains are identified in purple. \label{fig:coarsening}}
\end{figure}

The typical evolution of a sample is shown on Fig.~\ref{fig:coarsening},
for a temperature of $1230^\circ$C. The surface of the
barium-rich phase is represented on Fig.~\ref{fig:coarsening}, with
colors encoding the value of the local mean curvature (from dark green to
bright yellow for increasing curvature). The same cubic sub-volume of
linear size 250
$\mu\textrm{m}$
is represented at different times during an in-situ experiment, from 80
to 480 s. In this experiment, the volume fraction of the barium-rich
phase is measured to be $49 \pm 1\%$, so that it is in slight minority. 

An interconnected microstructure is observed for both phases. The
minority barium-rich phase is composed of smooth threads of fluid
connected together by thicker nodes. Such an interconnected
microstructure is often attributed to spinodal decomposition, however
our observations only start when the compositions of the phases are
fixed. A few disconnected barium-rich domains are observed as well (and
represented in purple in Fig.~\ref{fig:coarsening}), but they represent a
very small volume fraction of the phase (always less than 1 \% throughout
the experiment). No disconnected domains are ever observed for the
silica-rich phase. As time increases, we observe the coarsening of the
typical size of the microstructure (Fig.~\ref{fig:coarsening}), while the
morphology of the phases continues to look quite similar to the initial
state, up to a rescaling of the typical size. 

\begin{figure}
\centerline{\includegraphics[width=0.99\columnwidth]{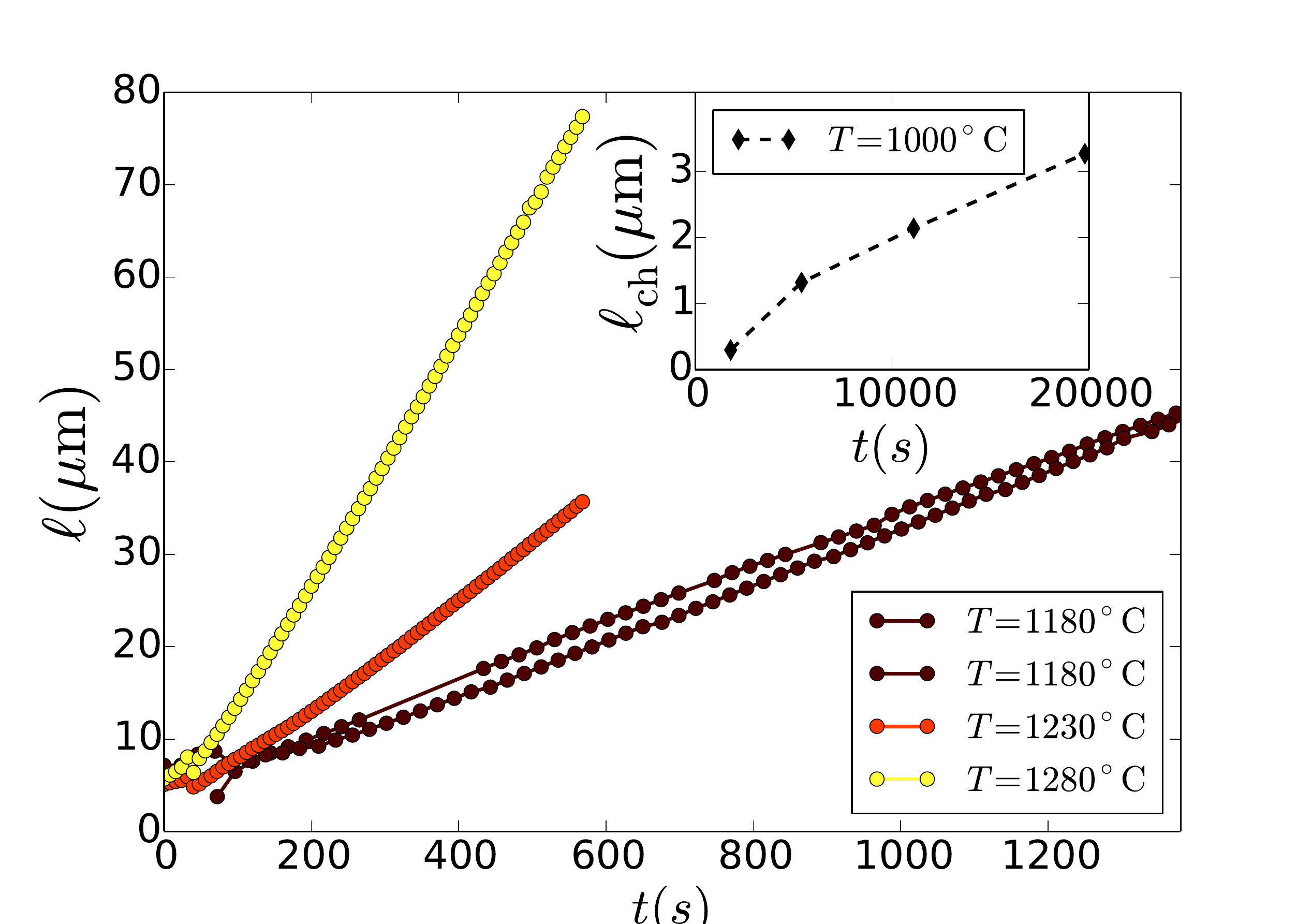}}
  \caption{Evolution of the characteristic length $\ell(t)$. Main axes: growth of $\ell$ measured in
in-situ tomography experiments at several temperatures. Inset: growth
of $\ell_\textrm{ch}$ measured post-mortem thanks to SEM imaging. \label{fig:lengths}}
\end{figure}

We measure the typical size $\ell$ of the barium-rich phase as its volume
$\mathcal{V}$
to surface $\mathcal{S}$ ratio:
\begin{equation}
\ell = \frac{3\mathcal{V}}{\mathcal{S}},
\end{equation}
so that $\ell$ is equal to the radius for a sphere. The surface area is
measured from the binarized image thanks to local histograms of voxel
configurations, as proposed by Lang et al.~\cite{Lang2001}. The
segmentation process often results in surfaces with a small-scale
roughness, at a scale below the true spatial resolution of the
experiment. Therefore, the surface is smoothened with a Gaussian filter
of standard deviation of 1 pixel before the surface area is computed.
This smoothing step reduces the sensitivity of $\ell$ to the measure and
segmentation noise. We have compared $\ell$ to other measurements of the
average local scale derived from the correlation function of the
binarized image, or the chord length distribution~\cite{Torquato1993,
Levitz2007, Levitz1991}. All quantities resulted in similar
results~\cite{Bouttesphd} (up to constant prefactors of order one, coming
from the different definitions).

\begin{figure}
\centerline{\includegraphics[width=0.99\columnwidth]{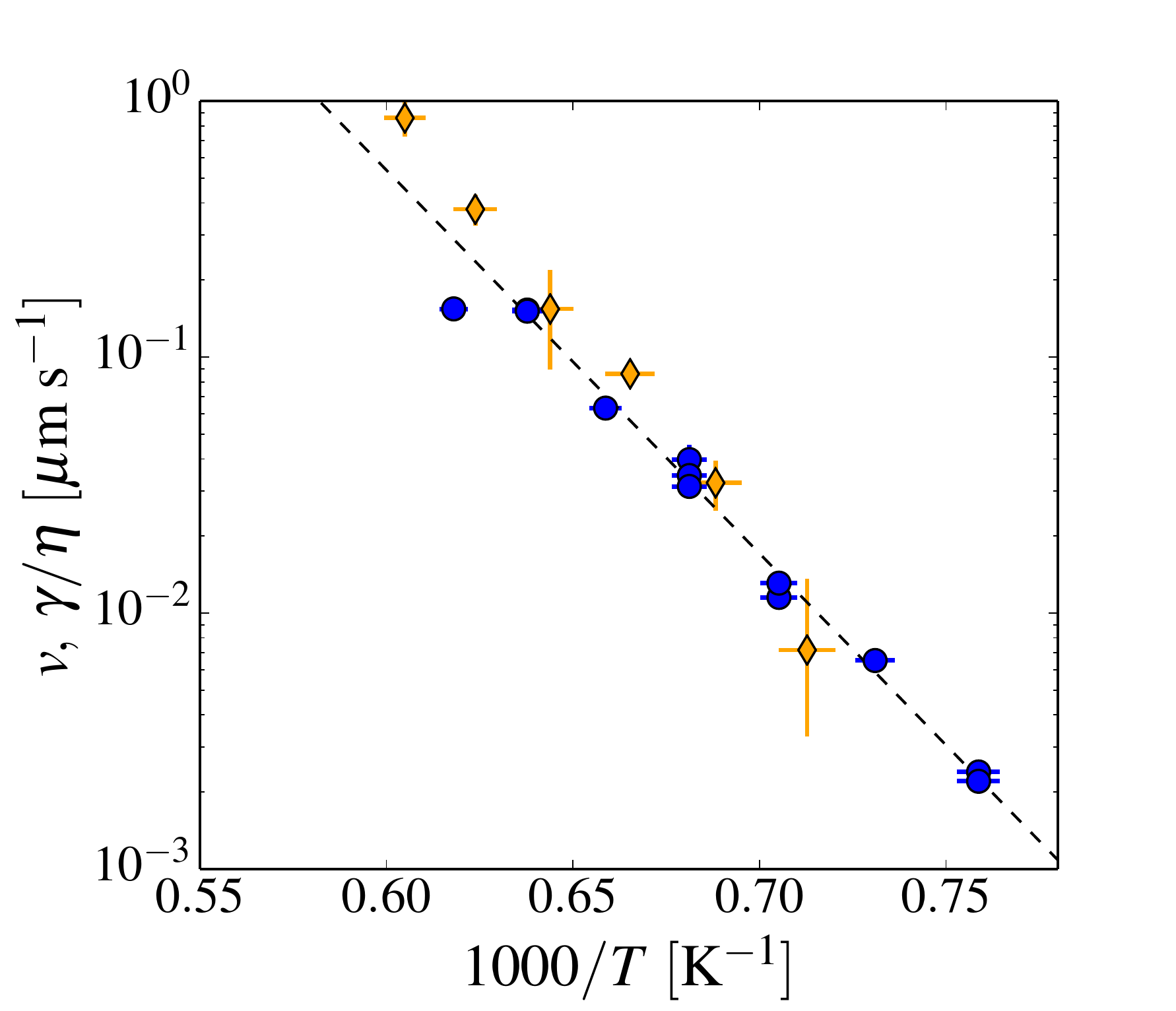}}
  \caption{Compararison between experimental and theoretical coarsening
rate, plotted in an Arrhenian diagram. Circles: linear growth rates
$v=\dot{\ell}$
measured in experiments at temperatures from 1030 to $1330^\circ$C.
Diamonds: ratio $\gamma / \eta$, determined using the value of $\gamma$
obtained numerically (see Appendix) and measured viscosities.
\label{fig:arrhenius}}
\end{figure}

The evolution of $\ell$ with time is shown on
Fig.~\ref{fig:lengths} for experiments realized at different
temperatures, from $1180^\circ$C to $1280^\circ$C. A linear evolution
with time is observed, up to typical sizes as high as 80~$\mu
\textrm{m}$. Since the initial microstructure has a size of about 5 $\mu
\textrm{m}$, the typical size is multiplied by more than one order of
magnitude during the duration of an experiment. The coarsening rate
increases with temperature, and several experiments performed at the same
temperature show that the value of the coarsening rate is well reproduced
(see Fig.~\ref{fig:lengths}). In order to characterize the
temperature-dependence of the coarsening rate, we have plotted
$\dot{\ell}$ in an Arrhenian diagram in Fig.~\ref{fig:arrhenius}, for
11 different experiments realized from $1030$ to $1330^\circ$C. We
observe an Arrhenian temperature dependence, with an activation energy of
the order of $300\,\textrm{kJ}.\textrm{mol}^{-1}$. Since the
immiscibility dome widens towards lower temperatures, experiments at
different temperatures in Fig.~\ref{fig:arrhenius} correspond to
different volume fractions of the barium-rich phase, from 39 to 52 $\%$.
Such variability does not seem to affect the Arrhenian behavior.

In order to test the validity of the linear coarsening regime at smaller
scales, we measured the typical size of interconnected phases for thermal
treatments at $1000^\circ$C of the quaternary glass with alumina, using
post-mortem SEM imaging. Since $\ell$ cannot be directly measured from 2D
images, we measured as a proxy for $\ell$ the average chord size
$\ell_\textrm{ch}$, the chord size distribution~\cite{Torquato1993} being
a stereological measure that should be identical in 2D and 3D for
statistically isotropic systems. As shown in the inset of
Fig.~\ref{fig:lengths}, a linear growth of $\ell_\textrm{ch}$ with time
is observed for this set of experiments as well, for values of
$\ell_\textrm{ch}$ ranging from 1 to 3 $\mu$m. The first data point,
corresponding to a scale of 300 nm, does not match as well the linear
fit. This slight discrepancy at short times and small scales is likely
to come from variations in the initial size of the microstructure, since
our post-mortem observations were realized on different samples. We
conclude that a linear coarsening can be observed at scales ranging from
1 to 80 $\mu$m, including scales below the resolution of our tomography
experiments.

\begin{figure}
\centerline{\includegraphics[width=0.99\columnwidth]{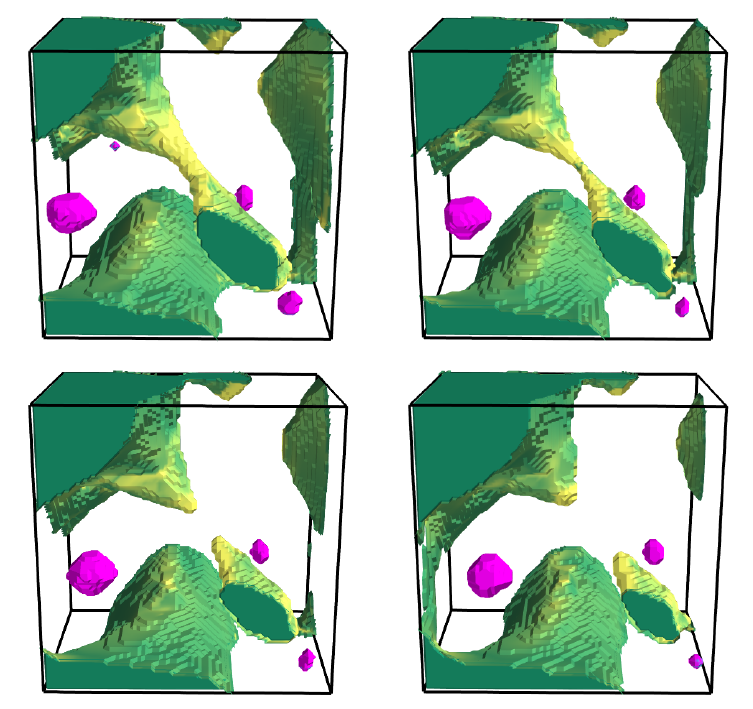}}
 \caption{Close-up visualization of a breaking liquid bridge. Surfaces of
the percolating domain are color-coded according to mean curvature, from
green to yellow. A few disconnected droplets are observed as well
(in purple). 
3-D images of linear size 140 microns, taken at times 408, 432, 440, and 464
s during an experiment at $1280^\circ$C.\label{fig:pinchoff}}
\end{figure}

The observed linear evolution of $\ell$ suggests that viscous coarsening
dominates over diffusive mechanisms. This may come as a surprise, since
diffusive coarsening has been the only mechanism reported so far in
silicate melts~\cite{Mazurin1985}. 3-D visualizations of the evolution of
interfaces (Fig.~\ref{fig:pinchoff}) unravel the local mechanisms
resulting in coarsening. Inside curved liquid bridges, Laplace pressure
causes fluid to flow from regions with a high mean curvature towards
regions with smaller curvatures. This results in a thinning of some
liquid bridges (Fig.~\ref{fig:pinchoff} (b)), up to the break-up of the
liquid filament (Fig.~\ref{fig:pinchoff} (c)). After the break-up, fluid
in the broken ends retracts because of capillarity and therefore flows in
adjacent regions (Fig.~\ref{fig:pinchoff} (c) and (d)), resulting in a
local coarsening. Local break-up events of viscous liquid bridges are
also known as \emph{pinch-off} in the fluid mechanics
literature~\cite{Eggers1997}. They were observed using confocal
microscopy in a colloid-polymer mixture by Aarts and
collaborators~\cite{Aarts2005}. Further evidence for the weak influence
of diffusion on coarsening is given by the fact that a few isolated
domains (displayed in purple in Fig.~\ref{fig:coarsening}) keep a
constant volume through time, so that evaporation can be considered
negligible up to the measurement precision. Visual observation also sheds
light on the origin of the few isolated domains, that break up from the
continuous phase during the retraction of long terminal
filaments~\cite{Bouttes2014} (see also Fig.~\ref{fig:gerris}).

Thanks to such direct observations of coarsening mechanisms, it is
possible to revisit the theoretical argument of Siggia~\cite{Siggia1979}
for deriving a
linear coarsening rate with time. One balances the dominant terms of the Navier-Stokes equation,
that are the Laplace pressure gradient and the viscous dissipation, when
inertia is negligible:
\begin{equation}
 \eta \frac{v}{\ell(t)^2} \sim \frac{p_{\mathrm{Laplace}}}{\ell}
\sim \frac{\gamma}{\ell(t)^2},
\end{equation}
with $\eta$ the largest viscosity of the two phases, $\gamma$ the
interfacial tension, $v$ a typical velocity of the fluid and the Laplace pressure $p_{\mathrm{Laplace}} \sim
\gamma / l$. Siggia argues that $v \sim \mathrm{d}\ell/\mathrm{d} t$, since
$\ell$ is the typical scale of fluid domains. 
In Fig.~\ref{fig:pinchoff},  $v$ would be a typical
velocity at which fluid flows from the bridge to the node, either when
the bridge is still connected or when the broken ends retract.
The growth
of $\ell$ in neighboring liquid bridges is due to the redistribution of
the fluid from the thinning, then breaking, liquid bridge. Replacing
$v$, one obtains the equation proposed by Siggia:
\begin{equation}
\ell(t) \sim \frac{\gamma}{\eta} t.
\label{eq:visc}
\end{equation}

In order to test whether Eq. (\ref{eq:visc}) describes well the
coarsening dynamics observed here, we have compared in
Fig.~\ref{fig:lengths} the evolution of the
linear prefactor with the evolution with temperature of the ratio $\gamma
/ \eta$. In Fig.~\ref{fig:arrhenius}, the linear
prefactors are plotted in an Arrhenian diagram, for temperatures ranging
from 1030 to $1330^\circ$C. As for the viscosity, the
kinetics of fluid rearrangement can be considered to be governed by the
largest viscosity. The estimated interfacial tension over the measured viscosity of the silica-rich phase is therefore superimposed on
Fig.~\ref{fig:arrhenius}. We observe a very good agreement between
$\dot{\ell}(t)$ and $\gamma / \eta$: the activation energies of the two
quantities are very close. Moreover, the proportionality coefficient
between $\dot{\ell}(t)$ and $\gamma / \eta$ seems to be close to one.
The good quantitative agreement between our measurements and Eq.
(\ref{eq:visc}) is a convincing evidence that viscous coarsening
is a mechanism present in phase-separated silicate melts.   

\section{Self-similarity of the microstructure \label{sec:scaling}}

\begin{figure}
\centerline{\includegraphics[width=0.99\columnwidth]{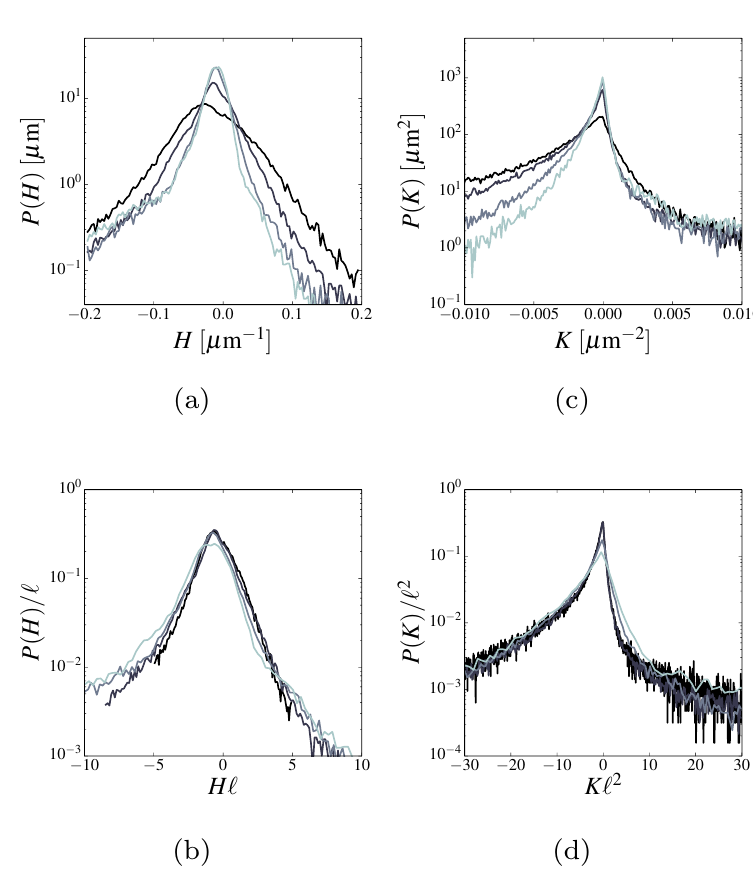}}
  \caption{Histograms of mean and Gaussian curvatures, for an experiment
at $1180^\circ$C, observed at thermal treatment times of 15, 28, 48 and
60 min, from dark blue to light blue. (a) Mean curvature $H$. (b)
Renormalized mean curvature $H\ell$. (c) Gaussian curvature $K$. (d)
Renormalized Gaussian curvature $K \ell^2$.\label{fig:curvature}}
\end{figure}

An important hypothesis underlying theoretical descriptions of coarsening
is that the geometry of the phases remains statistically the same up to a
rescaling of the typical size $\ell$ -- a hypothesis called
\emph{dynamic scaling}~\cite{Bray2003}. In particular, this argument is
used by Siggia~\cite{Siggia1979} to justify that $\dot{\ell}$ is equal to
the velocity induced by the Laplace pressure. 

Statistical characterizations of the domain geometry allow us to test the
dynamic scaling hypothesis. In order to characterize the geometry of the
system, we compute histograms of surface curvatures, for an experiment at
$1180^\circ$C. Distributions of curvatures are a common tool for a local
characterization of interfaces, in particular for coarsening
studies~\cite{Mendoza2003,Kammer2006}. Principal curvatures $\kappa_1$
and $\kappa_2$ are computed by fitting a quadric surface to the
neighbourhood of points lying on the interfacial
surface~\cite{Nishikawa2001, Lopez-Barron2009}. Mean and Gaussian
curvatures are defined respectively by
\begin{equation}
H = \frac{\kappa_1 + \kappa_2}{2} ;\quad K = \kappa_1\kappa_2.
\end{equation}
Histograms of $H$ and $K$ measured at times from 15 to 60 minutes are
plotted in Fig.~\ref{fig:curvature}. Histograms of curvatures get
narrower with time because of global coarsening, as the characteristic
length $\ell$ grows from 25 to 75 microns from the first to the last
image used to build the histograms. The range of curvature values
reflects the different possible morphologies of the interfacial surface,
with rounded tips ($K > 0$, $H < 0$), tubes ($|K| \ll \ell^{-2}$, $H <
0$) or saddles ($K < 0$). For both $H$ and $K$, the peak of histograms
correspond to tubular morphologies, and the tails to either tips of such
tubes, or junctions between them. Fig.~\ref{fig:curvature} (b) and (d)
shows the histograms of mean and Gaussian curvatures renormalized by the
characteristic length $\ell$, $H\ell$ and $K \ell^2$. A good collapse on a
master curve is observed both for mean and Gaussian curvature,
demonstrating that, statistically speaking, the morphology of the
phases does not change over time, except for their average size. The
validity of dynamic scaling means that it is possible to elaborate
morphologies that are statistically similar, but with sizes that are
different. Such property can be interesting for applications but also for
designing model materials.

Nevertheless, one should mention that deviations from dynamic scaling can
be observed at longer times, because of the progressive fragmentation of
disconnected droplets from the percolating low-viscosity phase. Such
domains can be observed in Fig.~\ref{fig:coarsening}
and~\ref{fig:pinchoff}. We have studied deviations from dynamic scaling
in a previous paper~\cite{Bouttes2014}. Recent work~\cite{Bouttes2015} has
shown that the fragmentation rate increases drastically when the volume
fraction of the low-density phase decreases. In the set of experiments
presented here, the volume fraction is large enough that a small
fragmentation rate hardly disrupts dynamic scaling for the duration of
experiments.

\section{Discussion \label{sec:discussion}}

\begin{figure}
\subfigure[]{
\centerline{\includegraphics[width=0.99\columnwidth]{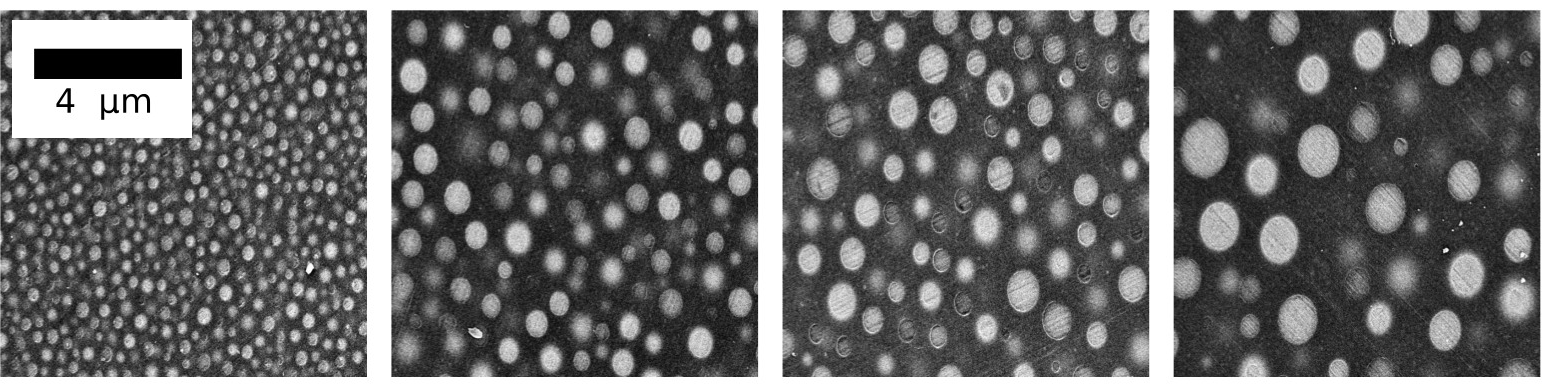}}
}
\subfigure[]{
\centerline{\includegraphics[width=0.7\columnwidth]{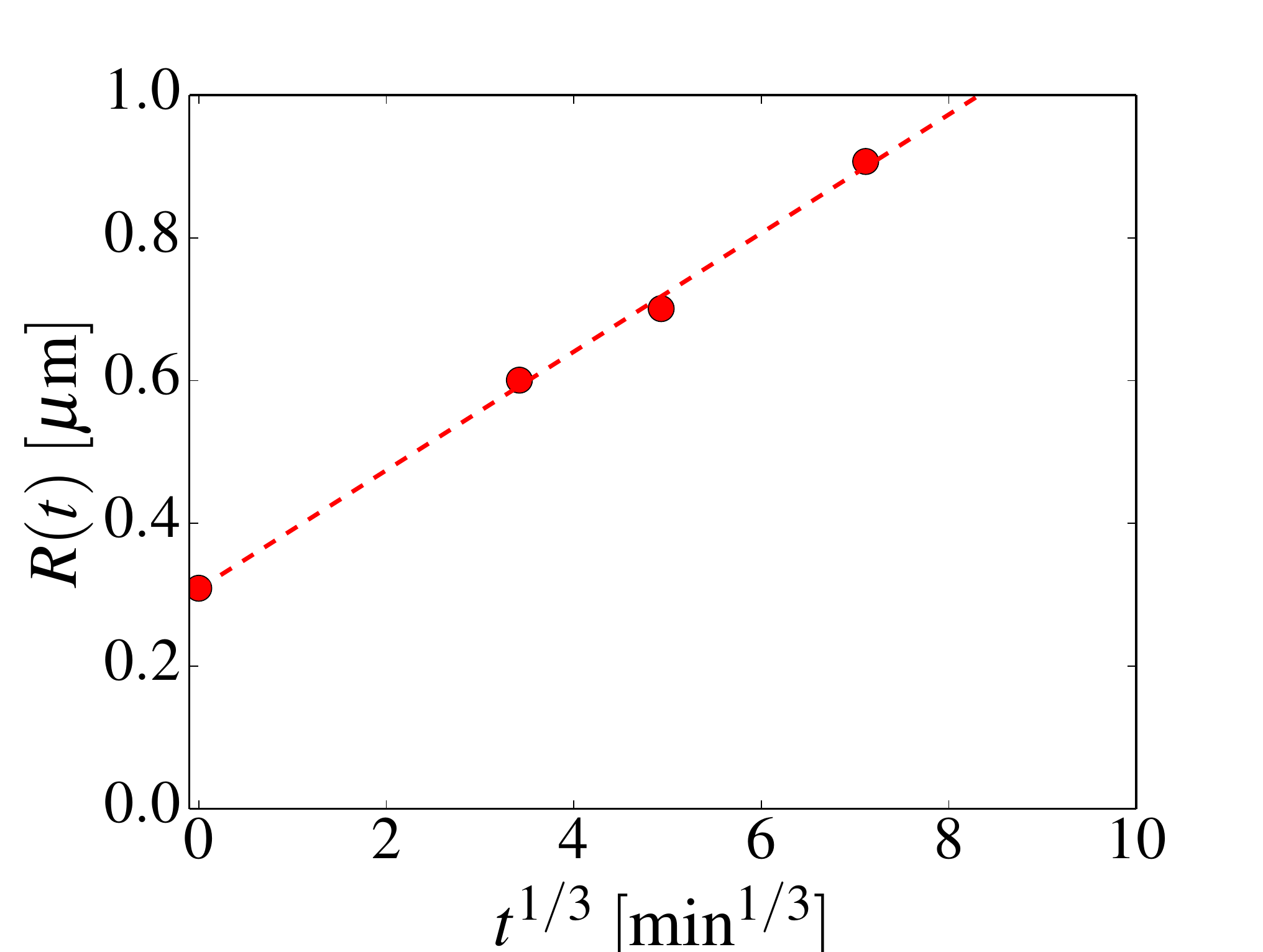}}
}
  \caption{(a) SEM images of coarsening for a droplet morphology, for
thermal treatments of 0, 20, 120 and 360 min at $1250^\circ$C (left to
right). (b) Evolution of the average droplet radius with $t^{1/3}$.
\label{fig:LSW}}
\end{figure}

Our tomography experiments demonstrate that viscous flow induced by
Laplace pressure is the prevailing coarsening mechanism over a wide range
of lengthscales. Then, why did not previous studies on phase separation
in glass-forming melts~\cite{Simmons1974,Dalmas2007} observe a
linear-with-time coarsening? In order to investigate the relative effects
of diffusion-induced and viscous-flow-induced coarsening, we carried on a
complementary set of experiments where diffusion was the sole coarsening
mechanism. To do so, we synthesized a composition on the same tie-line as
the ternary glass, but sufficiently off-centered towards the silica-rich
phase so that the composition lies inside the binodal domain (see
Fig.~\ref{fig:diagram}), and is characterized by a droplet morphology
(see Fig.~\ref{fig:LSW} (a)). After the quench from the stable region to
below the glass transition, a microstructure of barium-rich droplets
inside a silica-rich matrix is visible. In the absence of an
interconnected minority phase, the prevailing transport mechanism is
chemical diffusion through the boundary of the domains (droplets
coalescence because of Brownian motion~\cite{Tanaka1994} is bound to be
slower because of the micronic size of droplets and the very large
viscosity of the matrix). Thermal treatments were realized for cm-sized
glass cylinders at $1250^\circ$C for times ranging from 20 minutes to 6
hours. Polished surfaces cut through the samples were observed
post-mortem using SEM imaging. Typical SEM images are shown in
Fig.~\ref{fig:LSW} (a). The histogram of areas of the barium-rich
droplets was converted to an histogram of volumes of 3-D spheres, using a
classical stereologic computation~\cite{Russ2000}. The evolution of the
average domain size with time is shown in Fig.~\ref{fig:LSW} (b); it is
consistent with a $t^{1/3}$ evolution. In addition to the different
kinetic law, we note that the typical size of droplets is much smaller
than for the interconnected structure at a similar temperature (see
Fig.~\ref{fig:lengths}). This size discrepancy confirms that, when
topology allows it, viscous flow is a much more efficient mechanism than
diffusion for our system. 

For a diffusion-controlled coarsening, diffusion models predict that the
growth of $\ell$ evolves as~\cite{Bray1994}
\begin{equation}
\ell(t) \sim \left( \frac{\gamma D \Omega}{kT} t\right)^{\frac{1}{3}},
\label{eq:diff}
\end{equation}
where $D$ is the effective molecular diffusivity,
$\Omega$ a molecular volume, $k$ the Boltzmann constant and $T$ the
temperature. 
Fitting the size values of Fig.~\ref{fig:LSW} (b) with
Eq. (\ref{eq:diff}), we obtain an estimate for the product $D \Omega$.
Choosing the molecular volume $\Omega \simeq 10^{-29} \textrm{m}^3$ gives
$D \simeq 2.10^{-12}\,\textrm{m}^2\textrm{s}^{-1}$. The theoretical crossover length
$\ell^*$ between diffusive growth and hydrodynamic growth is commonly
estimated by equating the interfacial velocity in the two
regimes~\cite{Ahmad2012}, resulting in
\begin{equation}
\ell^* = \sqrt{\frac{\eta D \Omega}{kT}}.
\label{eq:crossover}
\end{equation}
For our system at $1250^\circ$C, the corresponding numerical values is
$\ell^* \simeq 10\,\textrm{nm}$, much below the initial microstructure size
in tomography experiments. The corresponding crossover time $t^* =
\frac{kT}{\gamma D \Omega} l^*$ would be of the order of 0.1 s. Observing only
the hydrodynamic regime at micronic scales was therefore to be expected.

In this respect, it is instructive to compare such orders of magnitude to
the ones of another study~\cite{Simmons1974} where both diffusivity and
viscosity were estimated. Simmons et al.~\cite{Simmons1974} measured the
evolution of the macroscopic viscosity resulting from coarsening in sodium
borosilicate glasses. Using the values obtained for $\eta$ and $D$
in~\cite{Simmons1974} and Eq. (\ref{eq:crossover}), one obtains a
crossover scale of a few microns and an hypothetical crossover time of
thousands of days. Observing the hydrodynamic regime requires $t^*$ to be
accessible in an experiment, and hence the viscosity of the glass not to
be too large. Since most previous studies in glass-forming melts were
realized slightly above the glass transition, the viscosity was always
very high ($> 10^{10}\,\textrm{Pa}.\textrm{s}$) and the hydrodynamic
regime was out of reach. An interesting perspective of our study would be
the investigation of the crossover of the diffusive and hydrodynamic
regimes. Such investigations would require to control the critical
temperature so that the viscosity just below $T_c$ would result in a
crossover time well resolved experimentally.

Let us finally note that simple models relating the viscosity and the
diffusivity, such as the Stokes-Einstein relation, or the Eyring
model~\cite{Eyring1936} for silicate glasses, result in a crossover
length $l^*$ of the same order as a typical molecular size.
Observations of diffusive coarsening over larger
scales~\cite{Dalmas2007,Simmons1974} therefore hint at more complex
diffusive mechanisms that are decoupled from viscosity -- a decoupling
that is well known for silicate glasses close to their glass
transition~\cite{Mungall2002, Liang1996} and that has been recently
discussed for metallic glasses as well~\cite{Zollmer2003}.

\section{Conclusions}

Using in-situ microtomography, we have shown that viscous flow prevails
as the dominant coarsening mechanism in phase-separated silicate melts,
for temperatures far above the glass transition. A linear growth of the
typical size has been observed for a large range of scales, from 1 to 80
$\mu$m. We have verified that the morphology remains statistically
self-similar during coarsening, up to a rescaling by the characteristic
size. Whether these mechanisms apply to other inorganic materials, such
as phase-separated bulk metallic glasses~\cite{He2013, Mattern2009}, is
an open question of interest for the fundamental understanding of such
materials.

The transition between the diffusive and viscous regime is an interesting
question open by this work, that will require to understand which
diffusive mechanisms control the coarsening at small scales. Future work
will focus on smaller volume fractions on the less-viscous phase, for
which more fragmentation of droplets is observed. 

\section*{Acknowledgements}

The authors are much grateful to Sophie Schuller for lending them the
falling-sphere apparatus for the viscosity measurements. The authors also
acknowledge precious help from Yohann Bale, Erick Lamotte, Rudy Vetro and
Jean-Paul Valade. This work was supported by the French ANR program
``EDDAM" (ANR-11-BS09-027). Experiments were performed on beamline ID19
at ESRF in the framework of projects HD501, SC3724, MA1839 and MA1876.

\section*{Appendix: estimation of interfacial tension}

\begin{figure}
\centerline{\includegraphics[width=0.99\columnwidth]{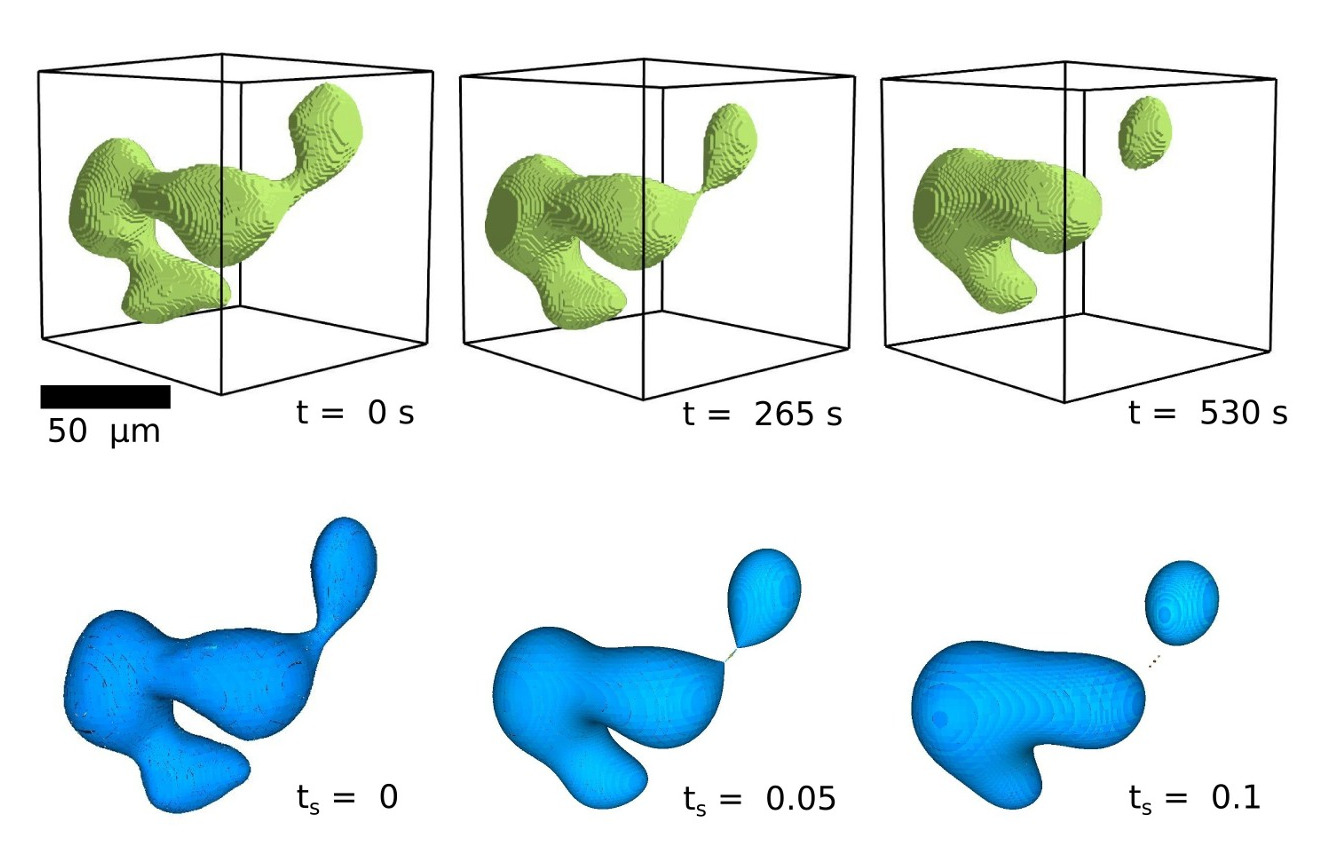}}
  \caption{Upper row: evolution of a single domain in an experiment at
$1180^\circ$C. The size of the square box is 190 $\mu$m. Lower row:
evolution of the same domain with the Gerris code, with a linear factor
between numerical and experimental times chosen to optimize the match
between geometries. \label{fig:gerris}}
\end{figure}

In order to estimate the interfacial tension between the two phases, we
performed numerical simulations with the Gerris
software~\cite{Popinet2003, Popinet2009}, a finite-volume solver of
Navier-Stokes equations. As an initial condition, a mesh was built from
the segmentation of a single domain in an experiment at $1180^\circ$C
(see Fig.~\ref{fig:gerris}). Physical parameters of the simulation were
set to one, including the size of the box, the interfacial tension
between the phases and the viscosity of the most viscous phase. The
viscosity of the least viscous phase was set to $10^{-2}$, ensuring a
high viscosity ratio between the phases. The viscosity ratio in the
simulation is not as large as the experimental one, since too large a
discrepancy between the two viscosities leads to prohibitive
computational times. Such parameters ensure that the Reynolds number is
smaller than 0.1. For Reynolds numbers much smaller than 1, in the Stokes
flow regime, inertia is negligible and time is only a parameter, with
boundary conditions determining completely the further evolution of the
system.

Fig.~\ref{fig:gerris} shows the evolution of the initial condition in the
experiment, as well as in the simulation of Navier-Stokes equations. A
striking similarity is observed between the experimental and numerical
geometries. In particular, the domain breaks up at the same place, while
the lower end of the domain retracts in both cases. Such observations
confirm that hydrodynamic flow is the mechanism responsible for the
observed coarsening. Moreover, the proportionality factor needed to
adjust experimental and non-dimensional numerical times gives the value
of the interfacial tension in the experiment, that is found to be $\gamma
= 10^{-2} \textrm{J}.\textrm{m}^{-2}$. We suppose that the interfacial
tension does not vary much with temperature, which is a reasonable
assumption compared to the much greater variations of the viscosities
with temperature.

\end{document}